\documentclass[aps,prl,twocolumn,superscriptaddress,asmsymb,amsmath]{revtex4-1}
\usepackage{graphicx}% Include figure files
\usepackage[utf8]{inputenc}
\usepackage{hyperref}
\usepackage{blindtext}
\usepackage{braket}
\usepackage{bm}% bold math
\usepackage{siunitx}
\usepackage{color}

\renewcommand{\vec}[1]{\mathbf{#1}}
\newcommand{\abs}[1]{\left|#1\right|}
\renewcommand{\k}{\vec{k}}
\newcommand{\q}{\vec{q}}
\newcommand{\p}{\vec{p}}

\begin{document}

\title{Phase transitions of polariton condensate in 2D Dirac materials}

\author{Ki Hoon Lee}
\affiliation{Center for Correlated Electron Systems, Institute for Basic Science (IBS), Seoul National University, Seoul 08826, Korea}
\affiliation{Department of Physics and Astronomy, Seoul National University, Seoul 08826, Korea}
\author{Changhee Lee}
\affiliation{Department of Physics and Astronomy, Seoul National University, Seoul 08826, Korea}
\author{Hongki Min}
\email{hmin@snu.ac.kr}
\affiliation{Department of Physics and Astronomy, Seoul National University, Seoul 08826, Korea}
\author{Suk Bum Chung}
\email{chung.sukbum@gmail.com}
\affiliation{Center for Correlated Electron Systems, Institute for Basic Science (IBS), Seoul National University, Seoul 08826, Korea}
\affiliation{Department of Physics and Astronomy, Seoul National University, Seoul 08826, Korea}
\affiliation{Department of Physics, University of Seoul, Seoul 02504, Korea}
%\author{Kihoon Lee$^{1,2}$, Changhee Lee$^1$, Hongki Min$^1$, Suk Bum Chung$^{1,2,3}$}
%\affiliation{$^1$ Department of Physics and Astronomy, Seoul National University, Seoul 08826, Korea}
%\affiliation{$^2$ Center for Correlated Electron Systems, Institute for Basic Science (IBS), Seoul National University, Seoul 08826, Korea}
%\affiliation{$^3$ Department of Physics, University of Seoul, Seoul 02504, Korea}
%\author{us}
\begin{abstract}
For the quantum well in an optical microcavity, the interplay of the Coulomb interaction and the electron-photon (e-ph) coupling can lead to %the emergence of bosonic quasiparticles consisting of the exciton and the cavity photon 
the %superpositions 
hybridizations of the exciton and the cavity photon 
known as polaritons, which can form the Bose-Einstein condensate above a threshold density. Additional physics due to the nontrivial Berry phase comes into play when the quantum well consists of the gapped two-dimensional (2D) Dirac material such as the transition metal dichalcogenide (TMDC) MoS$_2$ or WSe$_2$. Specifically, in forming the polariton, the e-ph coupling from the optical selection rule due to the Berry phase can compete against the Coulomb electron-electron (e-e) interaction. We find that this competition gives rise to a rich phase diagram for the polariton condensate involving both topological and symmetry breaking phase transitions, with the former giving rise to the quantum anomalous Hall and the quantum spin Hall phases. %depending on the excitation density.
\end{abstract}
\maketitle

%\begin{itemize}
%\item why polariton condensate?
%\item why TMDC? (1) Coulomb qualitatively important so Floquet not sufficient
%\end{itemize}

Monolayers of TMDC, such as MoS$_2$ and WSe$_2$, have attracted widespread interest in recent years as a semiconductor analogue of graphene. Like graphene, they are atomically thin, 2D materials with high mobility \cite{radisavljevic_mobility_2011}, with %Also, their D$_{3\textrm{h}}$ crystalline symmetry gives rise to 
the band extrema occurring at the Brillouin zone corners $K$ and $K'$ due to their D$_{3\textrm{h}}$ crystalline symmetry. %The most crucial similarity is the fact that we have Berry curvature concentrated at $K$ and $K'$ giving us the $\pm \pi$-Berry phase associated with each valley, as these extrema can be very well described by the Dirac Hamiltonian. 
Most crucially, these band extrema can be described very well by the Dirac Hamiltonian, which allows us to associate the $\pm \pi$-Berry phase with each valley, but %However, unlike the case of the graphene, this Dirac Hamiltonian is massive 
with direct band gaps at $K$ and $K'$ \cite{xiao_coupled_2012} unlike graphene. %Due to the broken inversion symmetry and spin-orbit coupling, the band edge at $K$ and $K'$  has no spin degeneracy, in stark contrast to the 4-fold degeneracy at the Dirac point for graphene.

Given %that the valley Berry phase originates from 
the band structure origin of the valley Berry phase, we may ask whether and how it may be affected by the %electron-electron 
e-e interaction and the %electron-photon 
e-ph coupling. It has been %much 
discussed recently that the superconductivity, %which %can be regarded as 
%arises from 
the condensation of %electron-electron 
e-e pairs %formed 
due to %electron-electron 
e-e interaction, of the doped TMDC is topologically non-trivial due to the valley Berry phase %at each valley 
\cite{yuan_tsc_2014, lu_tsc_2015, hsu_tsc_2017}. %A natural question that arises here is the possibility of the topologically non-trivial condensate of electron-hole pairs, {\it i.e.} excitons, 
Therefore, given the weak screening of the Coulomb interaction and the consequent strong binding of electron-hole pairs, {\it i.e.} excitons, that %has been discussed theoretically \cite{qiu_excitonoptic_2013} and observed experimentally \cite{ugeda_excitonoptic_2014}
is known to occur \cite{qiu_excitonoptic_2013, ugeda_excitonoptic_2014} in TMDC, it is natural to ask whether the condensate of excitons can also be topologically non-trivial. Meanwhile, the $\pi$-Berry phase %also greatly affects its 
has a well-known effect on the e-ph coupling, namely the optical valley selection rule for the circularly polarized light \cite{xiao_coupled_2012, cao_valley_2012, zeng_valley_2012}. %Given that both the electron-electron interaction and the electron-photon coupling is required for the polariton condensation, it is natural to ask whether any topological phase transition would occur due to the $\pi$-Berry phase. 
The TMDC monolayers are most convenient to manipulate optically, possessing direct band gaps ($\sim$ 1.5 to 2 eV) lying within the visible spectrum \cite{mak_gap_2010, splendiani_gap_2010}. %One can naturally ask if there is any condensation process in the TMDC monolayer that is tunable by both the Coulomb interaction and the electron-photon coupling and, if so, how would it be affected by 
%The above considerations motivate us to study the effect of the $\pi$-Berry phase on the condensation of the optically excited excitons.

%Hence, our aim is 
The above considerations motivate us to study the condensation of polaritons, emergent bosons from %superpositions 
hybridizations of cavity photons and excitons. It is tunable by both the Coulomb e-e interaction and the e-ph coupling, the former for the exciton energetics and the latter for the photon-exciton hybridization. The recent years have seen increasing consensus that this condensation has been experimentally observed in various systems \cite{deng_polariton_2002, kasprzak_boseeinstein_2006, balili_polaritonbec_2007} %including TMDC
with progresses underway for TMDC \cite{liu_tmdcPolariton_2015, chen_tmdcValleyPolariton_2017}. %With polaritons, condensation at room temperature is a possibility, 
The room temperature polariton condensation may be possible, %as %the mass of a polariton is especially small \cite{baumberg_high_2008} 
due to 
%light-matter coupling gives us 
an especially small polariton mass from light-matter coupling \cite{baumberg_high_2008}. %much smaller than that of an exciton. 
%While the finite lifetime of both cavity photons and excitons mean that polaritons do not exist in equilibrium, the same can also be said about the ultracold atoms in optical trap. 
%Taking the 
The finite lifetime of both cavity photons and excitons %into account, 
means that the polaritons %can 
exist in quasi-equilibrium. %like the ultracold atoms in optical trap.  
%Indeed, for %the condition analogous to the ultracold atom BEC, namely 
Despite the quasi-equilibrium nature,
in the case of the polariton lifetime  %being 
much longer than the thermalization time, substantial evidences of superfluidity, such as vortex formation \cite{lagoudakis_vortex_2008}, Goldstone modes \cite{utsunomiya_goldstone_2008}, and the Landau critical velocity \cite{amo_critical_2009}, have been observed. In this Letter, we will show %how, in the TMDC polariton condensation, the $\pi$-Berry phase gives rise to various phase transitions %coming from the competition between 
%due to the e-e interaction and the e-ph coupling. 
the $\pi$-Berry phase effects on the TMDC polariton condensate phase diagrams.

The polariton condensation in our gapped Dirac materials %the monolayer transition metal dichalcogenide such as MoS$_2$ and WTe$_2$ 
should be derived from the electrons with the Coulomb interaction coupled to coherent photons. Hence the Hamiltonian we consider would be %\cite{marchetti_thermodynamics_2006}
\begin{align}
\label{eq:total Hamiltonian}
\hat{H} & = \hat{H}_{\text{0}}+\hat{H}_{\text{e-e}}+\hat{H}_{\text{ph}}+\hat{H}_{\text{e-ph}}-\mu_{X}%\left(\hat{n}_{\text{ph}}+\frac{\hat{n}_{\text{c}}-\hat{n}_{\text{v}}}{2}\right)\label{eq:total Hamiltonian}\\
\hat{N}_{\rm tot},\\
\hat{H}_{\text{0}} & = \sum_{\tau=\pm}\sum_{\k}
\left[\begin{matrix}
	\hat{c}_{\tau,1,\k}^{\dagger} & \hat{c}_{\tau,2,\k}^{\dagger}
\end{matrix}\right]
%\left(\begin{matrix}m & \hbar v_{F}(k_{x}-ik_{y})\\
%\hbar v_{F}(k_{x}+ik_{y}) & -m
%\end{matrix}\right)
 {\bf d}_\tau^{(0)} (\k)\cdot {\bm \sigma}
 \left[
 \begin{matrix} 
 	\hat{c}_{\tau,1,\k}\\
 	\hat{c}_{\tau,2,\k}
\end{matrix}\right],\nonumber\\
%{\bf d}(\k) &=& (\hbar v_F k_x, \hbar v_F k_y, m),\nonumber\\
\hat{H}_{\text{ph}} & = \hbar\omega_c \sum_I \left( \hat{a}_I^{\dagger}\hat{a}_I+\frac{1}{2}\right),\nonumber\\
\hat{H}_{\text{e-ph}} & = %\frac{ie}{\hbar c}
-\frac{1}{c}\vec{\hat{A}}\cdot \sum_{\tau=\pm}\sum_{\k}\sum_{i,j} {\bf J}_{ij}^{\tau} (\k) \hat{c}_{\tau,i,\k}^{\dagger}  \hat{c}_{\tau,j,\k},\nonumber\\
% & = \frac{1}{\sqrt{A}}\sum_{\k}\left(\begin{matrix}\hat{\psi}_{c,\k}^{\dagger} & \hat{\psi}_{\k,v}^{\dagger}\end{matrix}\right)\left(\begin{matrix}0 & a\Omega_{\k}\\
%a^{\dagger}\Omega_{\k}^{*} & 0
%\end{matrix}\right)\left(\begin{matrix}\hat{\psi}_{c,\k}\\
%\psi_{\k,v}
%\end{matrix}\right),\nonumber\\
\hat{H}_{\text{e-e}}\!&=\!\frac{1}{2S}\!\sum_{\tau,\tau'}\!\sum_{\k_{1},\k_{2},\q}\!\sum_{i,j} V(q) \hat{c}_{\tau,i,\k_1-\q}^{\dagger} \hat{c}_{\tau',j,\k_2+\q }^{\dagger} \hat{c}_{\tau',j,\k_2} \hat{c}_{\tau,i,\k_1},\nonumber
\end{align}
where %$\hat{c}_{1(2)}$ is the electron annihilation operator in the orbital basis, 
${\bm \sigma}$ represent the Pauli matrices, $I$ the photon polarization index, ${\bf d}^{(0)}_\tau(\k) \equiv (\tau \hbar v k_x, \hbar v k_y, E_{\textrm{gap}}/2)$, with $\tau = \pm$ being the valley index, %giving us the massive Dirac dispersion $\sqrt{E_{\textrm{gap}}^2/4 + (\hbar v k)^2}$, 
$\omega_c$ the cavity photon frequency and $V(\q)=\frac{2\pi e^{2}}{\epsilon q}$ %giving us 
the Coulomb interaction, with $\epsilon$ being the dielectric constant; %, $\hat{c}_{c(v)}$ means the annihilation operator for electron in band basis from $\hat{H}_{0}$. 
%we will discuss later the consequence of considering 
note that the exchange terms of the %electron-electron 
e-e interaction are in the orbital rather than the band basis %as in the previous studies 
\cite{marchetti_thermodynamics_2006, kamide_meanfield_2010, byrnes_bcs_2010}. 
Meanwhile, the first quantized current operator is given by $\vec{J}_{ij}^{\tau} (\k) = -e\partial_\k[{\bf d}^{(0)}_\tau(\k)\cdot {\bm \sigma}]_{ij}$ and the gauge field operator %and the photon creation/annihilation operator are related 
by $\vec{\hat{A}} = \sum_I \sqrt{2\pi c^2 \hbar / \epsilon S L_c \omega_c} (\mathbf{e}_I\hat{a}_I e^{-i\omega_c t} + \mathbf{e}_I^* \hat{a}^\dagger_I e^{i\omega_c t})$, where $\vec{e}_I$ is the photon polarization vector and $S$, $L_c$ the cavity area and length, respectively. 
%As for the operators, 
$\hat{c}_{1(2)}$ and $\hat{a}_I$ are the annihilation operators for the electron in the $L_z = 0$ ($L_z = 2\tau$) orbital and the photon with the polarization $I$, respectively. 
%We take each 
Each valley is taken to be completely spin-polarized with opposite spin polarization, {\it i.e.} $S_z = \tau/2$, %as 
due to the transition metal atomic spin-orbit coupling ${\bf L}\cdot{\bf S}$ removing the spin degeneracy in the $L_z = 2\tau$ orbital, and no intravalley spin-flip process is considered; hence, the dark excitons from intravalley spin-flip \cite{Echeverry_spinDark_2016} will not be considered. 
Lastly, $\hat{N}_{\rm tot} = \sum_I \hat{a}^\dagger_I\hat{a}_I + \hat{N}_{\text{ex}}$ is the total number of excitations, both photons and excitons, in the system %that is 
and tuned by the %total excitation 
chemical potential $\mu_{X}$. Since the number of exciton $\hat{N}_{\text{ex}}$ is %basically 
the number of electrons excited from the valence band to the conduction band, %its definition requires us to write 
the band basis for the electrons, 
%\begin{align}
$\sum_\alpha \left[W(\k)\right]_{i,\alpha} \hat{\psi}_{\alpha,\k} =\hat{c}_{i,\k}$ %,\nonumber\\
%{\bf d}(\k)\cdot {\bm \sigma} W(\k)=&\left[\!\frac{E_{\textrm{gap}}^2}{4}\!+\!(\hbar v_{F} k)^{2}\!\right]^\frac{1}{2}\!W(\k)  
%\sigma^z,
%\left[
%\begin{matrix}
%	1 & 0\\ 
%	0 & -1\end{matrix}
%\right], 
%\end{align}
%Annihilation and creation operators $a$ and $a^{\dagger}$ are for circularly polarized photon. 
which diagonalizes $\hat{H}_0$ of Eq. \eqref{eq:total Hamiltonian} with $\hat{\psi}_{c(v)}$ as the annihilation operator of electrons in the conduction (valence) band, can be convenient. %The transformation matrix $W({\bf k})$ makes explicit the Berry phase of $\hat{H}_{\rm 0}$. %; the transformation matrix $W(\k)$ carries the $\pi$ Berry phase of the Dirac Hamiltonian. 
This allows to identify the exciton number as $\hat{N}_\text{ex} \equiv \sum_{\tau,\k} \hat{n}_{{\rm ex},\k}^\tau$ where $\hat{n}_{{\rm ex},\k}^\tau \equiv (\hat{\psi}^{\dagger}_{\tau,c,\k} \hat{\psi}_{\tau, c,\k} + \hat{\psi}_{\tau,v,\k} \hat{\psi}^{\dagger}_{\tau,v,\k} )/2$. Physically, we are interested in the thermal quasi-equilibrium that is reached after the cooling of a population of hot polaritons initially introduced by a short laser pulse \cite{byrnes_exciton-polariton_2014}. %To consider the simplest case %where 
%for condensation, %would arise, 
For simplicity, 
we shall set the temperature %of our model system 
to be zero.

We use the BCS variational wave function for the polariton condensate %with translational invariance  
\cite{kamide_meanfield_2010, byrnes_bcs_2010}
\begin{equation}
\ket{\Psi \{\Lambda_\pm\}}\!=\!\mathcal{N}\!\prod_{I,\tau=\pm,\k}\!e^{\Lambda_{I} \hat{a}_{I}^{\dagger}} (u_{\tau ,\k}\!+\!v_{\tau, \k} \hat{\psi}^{\dagger}_{\tau,c,\k} \hat{\psi}_{\tau,v,\k}) \ket{0}
\label{EQ:WF}
\end{equation}
%\begin{equation}
%|\Psi (\lambda)\rangle = e^{\sqrt{S}\lambda a^{\dagger}-S\lambda^2/2} \prod_\k (u_\k + v_\k \hat{\psi}^{\dagger}_{c,\k} %\hat{\psi}_{v,\k})|0\rangle,
%= \mathcal{N} \exp\left[\sqrt{S}\lambda a^{\dagger} + \frac{u_\k} {v_\k}\hat{\psi}^{\dagger}_{c,\k} \hat{\psi}_{v,\k}\right]\left|0\right\rangle,
%\label{EQ:WF}
%\end{equation}
with $\mathcal{N} = e^{-\sum_{I=\pm} \Lambda_I^2/2}$ and $|u_{\tau, \k}|^2 + |v_{\tau ,\k}|^2 = 1$, where $I=\pm$ corresponds to the right (left) circularly polarization $\vec{ e}_\pm = (1,\pm i)/\sqrt{2}$ and $|0\rangle$ is the ground state of $\hat{H}_{\text{0}}$, in which photons are absent and all the valence (conduction) band states are occupied (vacant). %The condition $|u_{\tau, k}|^2 + |v_{\tau ,\k}|^2 = 1$ makes the exciton component BCS-like, and the 
In this wave function, the photon component gives the coherent state with the number of photons $\langle\hat{a}_I^\dagger\hat{a}_I \rangle = \Lambda_I^2$ and of excitons %while the exciton number expectation value is given by 
$\langle\hat{N}_{\tau,\text{ex}}\rangle = \sum_\k |v_{\tau, \k}|^2$. 
%Eq.\eqref{EQ:WF} is equivalent to the ground state of Eq.\eqref{eq:total Hamiltonian} obtained through the mean-field approximation, as we are solving for 
To determine $\Lambda_\pm, u_{\tau,\k}, v_{\tau,\k}$ that minimize $\bra{\Psi \{\Lambda_\pm\}}\hat{H} \ket{\Psi \{\Lambda_\pm\}}$,
%Electron-electron interaction is transformed into mean-field form within Hartree-Fock approximation. 
%That means we are going %beyond applying the mean-field approximation 
%to apply 
we obtain the mean-field self-consistency condition not only for the %electron-electron 
e-e interaction through $\hat{H}^{\text{MF}}_{\text{e-e}} = \sum_{\tau,i,j,\k}\tilde{\Delta}_{\tau;ij}(\k) \hat{c}_{\tau,i,\k}^{\dagger} \hat{c}_{\tau,j,\k}$ where \footnote{The no-photon ground state value in the Fock potential is subtracted off so that %so that, with our band parameters, 
$\hat{c}_{\k}$'s can be treated as the non-interacting quasiparticles with our band parameters} %In the second line, %under an assumption that the 
\begin{equation}
\tilde{\Delta}_{\tau;ij}(\k)\!=\!-\frac{1}{S}\sum_{\p}V(\k\!-\!\p)\left.\left\langle \hat{c}_{\tau,j,\p}^{\dagger} \hat{c}_{\tau,i,\p}\right\rangle\right\vert^{\mu_X, \Lambda_\tau}_{\mu_X=0, \Lambda_\tau=0},
\label{eq:sceq_Fock}
\end{equation}
%where we are restricting ourselves to the exchange part ${\bf q} = \k - \mathbf{k'}$ and the Hartree terms $V(\mathbf{q}=0)$ are dropped to ensure charge neutrality and  
%note that we are 
%by subtracting off its 
%translational symmetry remains
%unbroken, 
%we assume the translational invariance, {\it i.e.} $\left\langle \hat{c}_{\mathbf{k_{1}-q},i}^{\dagger} \hat{c}_{\k_{2},j}\right\rangle =\delta_{\k_{1}-\mathbf{q},\mathbf{k_{2}}}\left\langle \hat{c}_{\mathbf{k_{2}},i}^{\dagger} \hat{c}_{\k_{2},j}\right\rangle $. 
%is used. 
but also for $\hat{H}_{\text{ph}}+\hat{H}_{\text{e-ph}}$, by which $\Lambda_I$'s are determined. %to obtain the self-consistency condition for the number of photons $\Lambda$. %as well. 
%Along with Eq.\ref{eq:sceq_lambda}, Eq(\ref{eq:sceq_Fock}) and Eq(\ref{eq:sceq_Num}) constitute the set of equations we solve in a self-consistent manner.
To obtain the latter condition, we apply rotating wave approximation on %electron-photon 
e-ph coupling 
%\begin{equation}
$\hat{H}_{\text{e-ph}} = \frac{1}{\sqrt{S}}\sum_{\k,I,\tau} g_\k^{I,\tau}\hat{a}_I \hat{\psi}^{\dagger}_{\tau,c,\k} \hat{\psi}_{\tau,v,\k} + {\rm h.c.}$,
%\end{equation}
where 
%\begin{equation*}
$g^{I,\tau}_{\k}=\sqrt{\frac{\hbar^3}{2\omega_c\epsilon S L_{c}}}%[W^\dagger(\k)]_{c,i}[\mathbf{e}_{R}\cdot {\bf J}(\k)]_{i,j} [W(\k)]_{j,v} 
\bra{c} \mathbf{e}_I \cdot {\bf J}^{(\tau)}(\k) \ket{v}$
%\end{equation*}
is the %electron-photon coupling 
e-ph strength, %from which we obtain
which gives us \cite{byrnes_bcs_2010}
\begin{equation}
\Lambda_I =-\frac{1}{\hbar \omega_c-\mu_{X}}\frac{1}{\sqrt{S}}\sum_{\tau,\k}\left(g^{I,\tau}_{\k}\right)^{*}\braket{ \hat{\psi}_{\tau,v,\k}^{\dagger}\hat{\psi}_{\tau,c,\k}}.
\label{eq:sceq_lambda}
\end{equation} 
%When we use 
%For the photons in the circular polarization basis, %for the photons, {\it i.e.} $I=\pm$, 
%the %electron-photon 
%e-ph coupling strength $g^{I,\tau}_\k = g_0 \delta_{I,\tau} + O(k^2)$ gives us the 
The optical valley selection rule \cite{xiao_coupled_2012} gives us %in other words, $g^{I,\tau}_\k$ has 
the $s$-wave symmetry for %$I=\tau$. 
the e-ph coupling, {\it i.e.} $g^{I,\tau}_\k = g_0 \delta_{I,\tau} + O(k^2)$, where $I$ is the photon circular polarization index.

From the self-consistency conditions of Eqs.~\eqref{eq:sceq_Fock} and \eqref{eq:sceq_lambda}, we find that there exists the competition between the e-e interaction and the %electron-photon 
e-ph coupling in the polariton condensation in the Dirac material. %arising from its 
%The 
%$\pi$-Berry phase. %of the Dirac material should play a crucial role in determining the symmetry of the electron-hole pair forming an exciton. 
%To see this, it is best to consider only the $\tau = +$ valley with the circularly polarized light, $\vec{ e}_+ = (1,i)/\sqrt{2}$. 
We first note that, %due to the optical valley selection rule arising from %a known consequence of 
%the $\pi$ Berry phase, %leads to the electron-photon coupling favoring the $s$-wave pairing for excitons. 
the absorption of the right (left) circularly polarized photon creates %implies creation of 
the $s$-wave, {\it i.e.} isotropic, exciton at the %$+$ ($-$) 
$\pm$ valley, %from the absorption of the right circularly polarized photon. 
%This is because %in Eq.\eqref{eq:sceq_lambda}, for a given amount of photon $\Lambda$,  it is favorable for $g_\k$ and the exciton $\left\langle \hat{\psi}_{c,\k}^{\dagger}\hat{\psi}_{v,\k}\right\rangle$ to have the same symmetry, and the 
%$g^{I,\tau}_\k = g_0 \delta_{I,\tau} + O(k^2)$ has the $s$-wave symmetry near the valley points for $I=\tau$, {\it i.e.} %in the sense that %between the $\vec{e}_+$ polarized photons and the $\tau = +$ valley found in the optical valley selection rule is $s$-wave in the sense that 
%it does not vanish as $k \to 0$ 
%\cite{xiao_coupled_2012}, and  
as $\Lambda_I$ in Eq.~\eqref{eq:sceq_lambda} %favors the same symmetry for the %electron-photon 
is maximized when the 
e-ph coupling $g^{I,\tau}_\k$ and the exciton correlation $\left\langle \hat{\psi}_{\tau,c,\k}^{\dagger}\hat{\psi}_{\tau,v,\k}\right\rangle$ %for a non-vanishing $\Lambda_I$.  
are in the same symmetry. 
On the other hand, the %electron-electron 
e-e interaction may not favor the $s$-wave exciton when we examine $\hat{H}^{\text{MF}}_{\text{e-e}} %= \sum_{\tau,i,j,\k}\Delta_{\tau;ji}(\k) \hat{c}_{\tau,j,\k}^{\dagger} \hat{c}_{\tau,i,\k} 
= \sum_{\tau,\alpha,\beta,\k} \Delta_{\tau;\beta\alpha}(\k) \hat{\psi}_{\tau,\beta,\k}^{\dagger} \hat{\psi}_{\tau,\alpha,\k}$, %as can be seen from
given that
\begin{align}
%\begin{equation}
\Delta_{\tau;c,v}(\k)\!=&\!\sum_{i,j} [W\!\tilde{\Delta}\!W^\dagger]_{\tau;c,v} (\k)\!\approx\! 
%\approx 
\Delta_{\tau;c,v}^s (k)\!+\! %(\hat{k}_x - i\tau\hat{k}_y) 
e^{i\tau\phi_\k}\Delta_{\tau;c,v}^p (k),\nonumber\\
%\end{equation}
%\label{EQ:valleyChirality}
%\end{align}
%with 
%\begin{align}
\Delta_{\tau;c,v}^s (k)\!\approx&\!-\!\frac{1}{S}\!\cos^2 \frac{\theta_k}{2} \sum_\p\!V(|\k\!-\!\p|) \langle\hat{\psi}^{\dagger}_{\tau,v,\p} \hat{\psi}_{\tau,c,\p}\rangle \cos^2 \frac{\theta_p}{2},\nonumber\\
\Delta_{\tau;c,v}^p (k)\!\approx&\!-\!\frac{2}{S}\!\sin \theta_k \sum_\p\!V(|\k\!-\!\p|) %\langle\hat{\psi}^{\dagger}_{\tau,c,\p} \hat{\psi}_{\tau,c,\p}\!+\!\hat{\psi}_{\tau,v,\p} \hat{\psi}^{\dagger}_{\tau,v,\p}\rangle  
\langle \hat{n}^\tau_{{\rm ex},\p}\rangle\cos \theta_p, 
\label{EQ:exciton}
\end{align}
where $\tan \phi_\k \equiv k_y/k_x$, $\tan \theta_k \equiv \hbar v k / (E_{\textrm{gap}}/2)$. %and $n^\tau_{{\rm ex},\p} \equiv \langle\hat{\psi}^{\dagger}_{\tau,c,\p} \hat{\psi}_{\tau,c,\p}\!+\!\hat{\psi}_{\tau,v,\p} \hat{\psi}^{\dagger}_{\tau,v,\p}\rangle/2$ is the exciton number expectation value at $\p$. %(for this approximation, we take into account that the excitons are first created at small $k$). 
%As the $p$-wave component $\Delta_{\tau;c,v}^p$ arises from the $\tau\pi$ Berry phase, 
%That the 
The $p$-wave components $\Delta_{\tau;c,v}^p (k)$ arises from the $\tau\pi$ Berry phase, as can be seen both from the chiralities of the $p$-wave components for the two valleys being opposite and $\Delta_{\tau;c,v}^p (k)$ being proportional to $\sin \theta_k$, the integrated Berry curvature for momenta smaller than $k$, that vanishes linearly as $k \to 0$.  %through the transformation matrix $W(\k)$ between the orbital and the band basis. %We see from 
%and the larger $E_{\rm gap}$, which reduces the Berry curvature around the valley point, suppresses the $p$-wave component through the prefactor $\sin \theta_\k$ of Eq.~\eqref{EQ:exciton}.  
%Furthermore, we can infer 
We see from Eq. \eqref{EQ:exciton} that the Coulomb e-e interaction favors the $p$-wave ($s$-wave) exciton at the $\tau$ valley when the $\tau$-valley exciton density %expectation value  %$n^\tau_{\rm ex} \equiv \sum_\k \langle\hat{\psi}^{\dagger}_{\tau,c,\k} \hat{\psi}_{\tau,c,\k} + \hat{\psi}_{\tau,v,\k} \hat{\psi}^{\dagger}_{\tau,v,\k} \rangle/2S$ 
$\sum_\p \langle \hat{n}^\tau_{{\rm ex},\p} \rangle$ becomes sufficiently large (small) 
compared to the critical density set by the average Berry curvature. %the first term, which creates 
%the chiral $p$-wave exciton $\Delta_{\tau;c,v}^p$ %through %phase winding from 
%$(\hat{k}_x - i\hat{k}_y) \hat{\psi}^{\dagger}_{c,\k} \hat{\psi}_{v,\k}$, 
%is going to be dominant in that valley, as 
%One reason for this is that %This is because %, in Eq.~\eqref{EQ:exciton}, $\langle\hat{\psi}^{\dagger}_{\tau,c,\p} \hat{\psi}_{\tau,c,\p}  + \hat{\psi}_{\tau,v,\p} \hat{\psi}^{\dagger}_{\tau,v,\p}\rangle$ is finite for all ${\p}^2 \lesssim 4\pi N^\tau_{\rm ex}/ S$ while $\langle\hat{\psi}^{\dagger}_{\tau,v,\p} \hat{\psi}_{\tau,c,\p}\rangle$ is negligible unless ${\p}^2 \approx 4\pi N^\tau_{\rm ex}/ S$.   
%Moreover, we will see that when this 
%when $\langle\hat{\psi}^{\dagger}_{\tau,c,\p} \hat{\psi}_{\tau,c,\p}  + \hat{\psi}_{\tau,v,\p} \hat{\psi}^{\dagger}_{\tau,v,\p}\rangle = 2\langle\hat{\psi}^{\dagger}_{\tau,c,\p} \hat{\psi}_{\tau,c,\p}\rangle$ is maximized, our ground state ansatz Eq. \eqref{EQ:WF} tells us that 
%$|\langle\hat{\psi}^{\dagger}_{\tau,v,\p} \hat{\psi}_{\tau,c,\p}\rangle| = \langle\hat{\psi}^{\dagger}_{\tau,c,\p} \hat{\psi}_{\tau,c,\p}\rangle \sqrt{1-\langle\hat{\psi}^{\dagger}_{\tau,c,\p} \hat{\psi}_{\tau,c,\p}\rangle^2}$ will be minimized.   
We will show that when %we have the predominantly chiral $p$-wave %component of 
%symmetry for 
the exciton symmetry of the polariton condensate %becomes dominant 
is predominantly chiral $p$-wave 
in the $\tau$ valley, %has the effect of changing 
the Berry phase sign of $\tau$ valley changes in the mean-field Hamiltonian $\hat{H}^{\text{MF}} \equiv \hat{H}_0 + \frac{1}{\sqrt{S}}\sum_{I,\tau,\k} (\Lambda_I g^{I,\tau}_\k \hat{\psi}^{\dagger}_{\tau,c,\k}\hat{\psi}_{\tau,v,\k} + {\rm h.c.}) + \hat{H}^{\text{MF}}_{\text{e-e}} -\mu_X \hat{N}_\text{ex}$ from that of $\hat{H}_0$. 
While %this term violates the conservation of the total excitation number $n$, the same is true even for the original 
Eq.~\eqref{EQ:exciton} also indicates that the chiral $p$-wave excitons are due to a component of the %Coulomb 
e-e interaction %$\hat{H}_{\text{e-e}}$ 
that violates the $N_{\rm tot}$ conservation, %of the total excitation number $N_{\rm tot}$,  
%Given that this is a quantitatively significant term that gives rise to the competition between two orthogonal exciton pairing symmetries, its exclusion will erase out qualitative features of the physics of the Dirac material polariton condensation. 
%it should be noted  that %even with this absence of the $n$ conservation, 
%the mean-field solution still %does 
%represents a stationary solution with very small deviation from the expectation value, as can be seen from 
the $N_{\rm tot}$ fluctuation remains small, {\it i.e.} $\frac{\langle(\Delta \hat{N}_{\rm tot})^2\rangle}{N_{\rm tot}^2} = \frac{\Lambda^2 + \sum_\k \abs{u_\k}^2 \abs{v_\k}^2}{(\Lambda^2 + \sum_\k \abs{v_\k}^2)^2} \ll 1.$ 
%little deviation of $n$ from the expectation value. 
%Because of this, rather than the chemical potential $\mu_X$, 
%Consequently, we have taken $N_{\rm tot}$ rather than $\mu_X$ as our control parameter, as we may expect it to increase monotonically with the intensity of the initial laser pulse in an actual experiment; one way to implement this in the self-consistent calculation is to invert Eq.\eqref{eq:sceq_lambda} so as to obtain $\mu_X$ from the fixed value of $\Lambda$.
%Moreover, since the natural physical control parameter for this system would be the intensity of the initial laser beam, $N_{\rm tot}$ rather than $\mu_X$ may be more natural as the control parameter in our calculation, which would require inverting the self-consistency condition of Eq.~\eqref{eq:sceq_lambda} to obtain $\mu_X$ from the fixed value of $\Lambda$.

\begin{figure}[t]
\centering{ \includegraphics[width= \columnwidth]{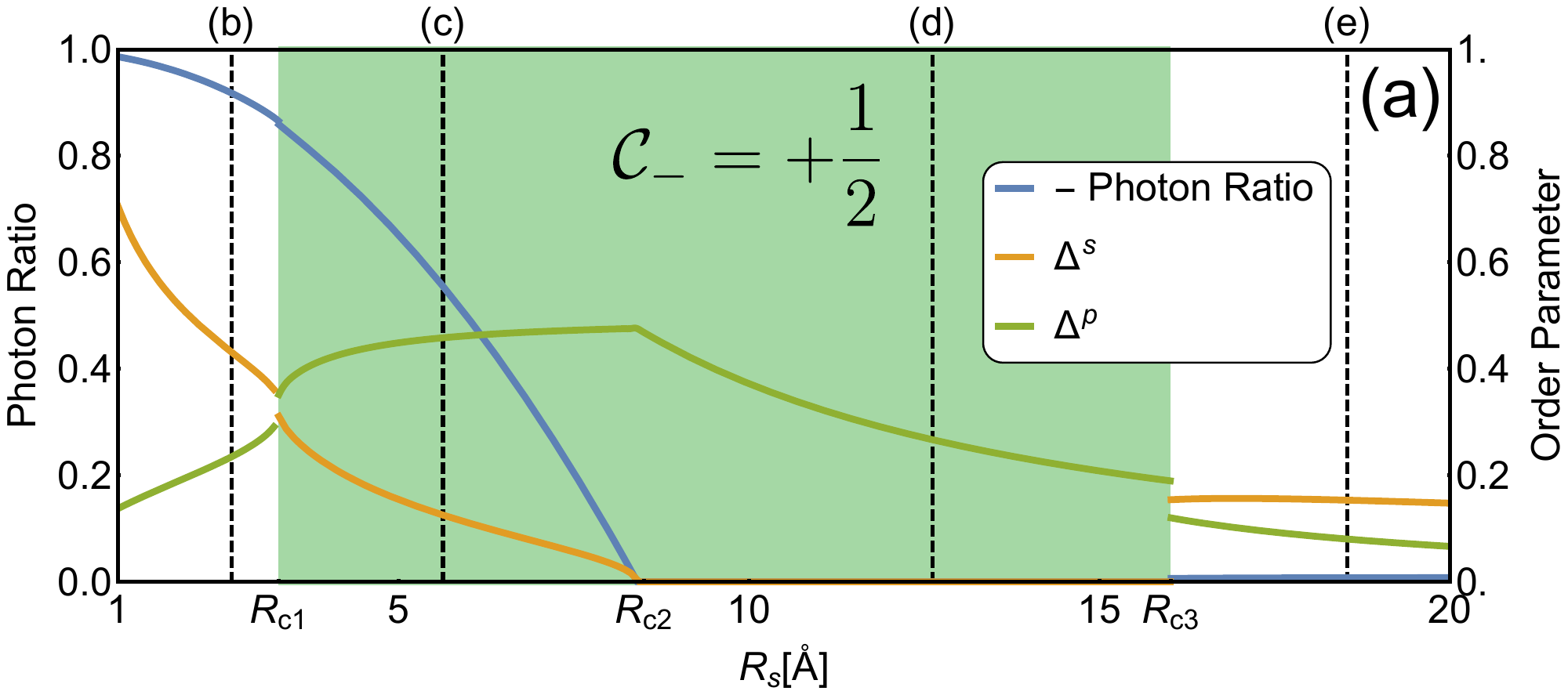}}\vspace{1em}
\centering{ \includegraphics[width= \columnwidth]{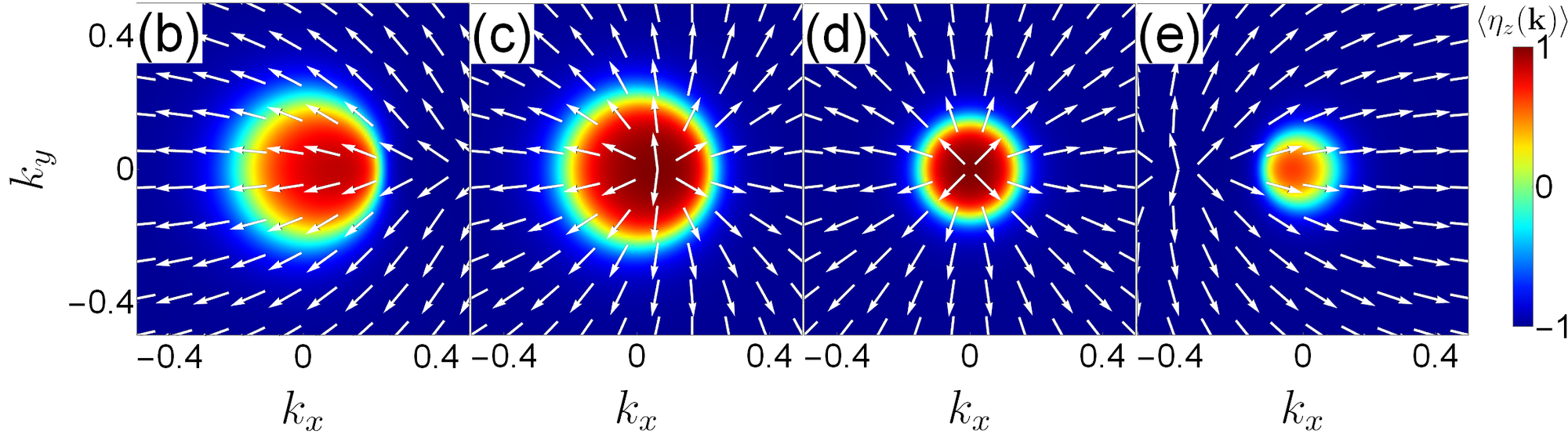}}

	\caption{(a) Photon fraction and mean-field band exciton gap parameters $\Delta^{s,p}$  %and $\mu_X$ 
		averaged over the momentum space as the functions of $R_s$ for the photon frequency $\hbar \omega_c = \SI{2.1}{eV}$, the dielectric constant $\epsilon = 10$, the Dirac velocity $\hbar v = \SI{3.7}{eV \AA}$, and the band gap of $E_{\textrm{gap}} = \SI{2.0}{eV}$. (b)-(e) Pseudo-spin textures at the $R_s$ values indicated in (a). Arrow represents %the direction of $\eta_x(\k) \hat{x} + \eta_y(\k) \hat{y}$ 
$\hat{\bm \eta}_\parallel$ and false color represents $\eta_z(\k)$; for convenience, we have plotted the $\tau = -$ valley coupled to $I=-$ photons.
}
	\label{FIG:phase1}
\end{figure}

%To illustrate 
The essence of the competition between the %Coulomb 
e-e interaction and the %electron-photon 
e-ph coupling %, it is instructive to first confine our problem to 
can emerge clearly from considering only a single valley, {\it i.e.} the $\tau = -$ valley coupled to the $I=-$ photons, %;  this to the TMDC polariton condensate with the initial laser pulse has the $I=+$ circular polarization. %with the circularly polarized light, $\vec{ e}_+ = (1,i)/\sqrt{2}$. 
%The numerical solution to the self-consistency equation for this single valley problem %the $\tau = +$ valley with the ${\bf e}_+$ polarization 
%shows 
revealing how the competition %between the e-ph coupling and the e-e interaction 
can give rise to the phase transition of our polariton condensate. Fig. \ref{FIG:phase1}(a) shows how the photon fraction $\Lambda^2 / (\Lambda^2 + \langle \hat{N}_\text{ex} \rangle)$ of the polariton condensate %, the chemical potential $\mu_X$, 
and %the gap of the quasiparticle spectrum from $\hat{H}^{\textrm{MF}}$; 
the exciton gap parameters $\Delta^{s,p}$ of Eq.\eqref{EQ:exciton} depend on the mean distance $R_s$ between excitations, the quantity that determines the total number of excitations $N_{\rm tot}$. %where two discontinuities indicates phase transitions. We emphasize that these 
%We see that the photon fraction plot %and $\mu_X$ 
%does not always decrease monotonically with $R_s$ and moreover has two discontinuities where the sign of the Berry phase is reversed; these features were not seen in the mean-field calculation of the polariton condensate in the quantum well system without the $\pi$ Berry phase, where only crossovers, not transitions, have been found between different phases. 
%The $\mathcal{C}=\pi$ Berry phase arises when the exciton symmetry is dominated by the chiral $p$-wave component, as this reversal of the Berry phase sign from its value for $\hat{H}_0$ occurs for $|\Delta^p| > |\Delta^s|$. %The crossings between $|\Delta^s|$ and $|\Delta^p|$ curves are discontinuous, with the discontinuity being smaller at the higher density (or the smaller $R_s$) transition.  
%For the transition at the higher density (or the smaller $R_s$), the discontinuity is smaller, with the quasiparticle gap nearly vanishing. 
%Another conspicuous feature seen 
%We also see 
%One conspicuous 
A %notable feature 
key feature here is that the %chiral
$p$-wave excitons are dark \cite{Ye_darkExciton_2014}, %`dark' in the sense that %the larger $|\Delta^p|$ suppresses the photon fraction. 
%the photon fraction %decreases with the increasing $|\Delta^p|$ and vanishes together with %of the photon fraction coincides with the vanishing of 
%vanishes for $\Delta^s=0$, %. This can be understood from Eq.~\eqref{eq:sceq_lambda} %, which %for a nearly constant $g_{\bf k}$, 
%gives us a vanishing 
which can be %inferred 
confirmed from 
%where 
$\Lambda$ vanishing in Eq.\eqref{eq:sceq_lambda} for %when we have 
the purely $p$-wave %symmetry %for 
%the  
%exciton correlation 
$\langle \hat{\psi}^{\dagger}_v \hat{\psi}_c \rangle$ %with the purely $p$-wave symmetry %with 
because the $s$-wave symmetry for the %electron-photon 
e-ph coupling, {\it i.e.} $g_\k \approx g_0 \delta_{I,\tau}$. %, leads to the vanishing photon fraction for $\Delta^s=0$.  
Since $\Delta^p$ arises solely from the %Coulomb 
e-e interaction, the higher-density discontinuous crossing of $|\Delta^s|$ and $|\Delta^p|$ curves %at the higher density %(or smaller $R_s$) 
in Fig.~\ref{FIG:phase1}(a) at $R_s = R_{c1}$ %where the discontinuous drop in the photon fraction also occurs, 
can be regarded as a consequence of the competition between the e-e interaction and the %electron-photon 
e-ph coupling. 
%Indeed, %for the lower density (or larger $R_s$) values in the $\mathcal{C}=\pi$ region, 
%Fig. \ref{FIG:phase1} (a) shows the vanishing of the photon fraction to coincide with the vanishing of $\Delta^s$, an ultimate illustration of this competition. %in is nearly constant, %shows that any non-$s$-wave 
%the ch excitons can be considered as `dark' with respect to the ${\bf e}_+$ photons, the vanishing of the photon fraction for a large range of $R_s$ between the two phase transitions indicate the purely chiral $p$-wave symmetry of excitons. 
%Hence we can regard the exciton symmetry in the $\mathcal{C} = \pi$ region to be largely dominated by the chiral $p$-wave component. 

%While we find the neither transitions to be continuous, 
The lower density %(or larger $R_s$) 
transition %between (d) and (e) 
in Fig.~\ref{FIG:phase1}(a) at $R_s = R_{c3}$ %does not significantly 
involves little %electron-photon 
e-ph coupling, and 
can be attributed to the competition between different components of the e-e interactions shown in Eq.\eqref{EQ:exciton},  
%Even when the photon fraction is negligible, the Coulomb interaction itself can favor the $s$-wave exciton if $N_{\rm ex}$ is small enough so that both $\Delta^s$ and $\Delta^p$ of Eq.\eqref{EQ:exciton} would arise primarily from the valley point ${\p} = 0$. 
%where the Coulomb interaction favors 
which favors the chiral $p$-wave exciton for the large $\sum_\k \langle \hat{n}^\tau_{{\rm ex},\k}\rangle$ (or small $R_s$) and the $s$-wave exciton for the small $\sum_\k \langle \hat{n}^\tau_{{\rm ex},\k}\rangle$ (or large $R_s$).  
%That this lower density transition is qualitatively more discontinuous than the higher density transition can be shown from 
As shown in Fig. \ref{FIG:phase1}(b)-(e), %both the higher and the lower density transition 
the transitions at both $R_s = R_{c1}$ and $R_s = R_{c3}$ can be %effectively 
illustrated using the pseudo-spin texture defined from the parametrization of the mean-field Hamiltonian: $\sum_{\k,\alpha,\beta} \hat{\psi}^\dagger_{\alpha,\k} [{\bm \sigma} \cdot {\bm \eta}(\k)]_{\alpha\beta} \hat{\psi}_{\beta,\k} \equiv \hat{H}^\textrm{MF}$.  
%The main qualitative difference with the 
One can see that both transitions involve %the singular point of $\hat{\bm \eta}_\parallel$, the direction of the $xy$-plane projection of ${\bm \eta}$, moving in and out of the $\eta_z > 0$ region. 
the (dis)appearance of the skyrmion texture in the $\hat{\bm \eta}$ configuration, which requires in the $\eta_z > 0$ region the singularity of $\hat{\bm \eta}_\parallel \equiv \hat{\bm \eta} - (\hat{\bm \eta} \cdot {\bf \hat{z}}){\bf \hat{z}}$, %the direction of the $xy$-plane projection of ${\bm \eta}$, 
while $\hat{\bm \eta}=-\hat{z}$ as $k \rightarrow \infty$. However, Fig. \ref{FIG:phase1} (d) and (e) show that, for the %lower density 
$R_s = R_{c3}$ transition, the $\hat{\bm \eta}_\parallel$ singularity jumps from $\k=0$ to the $\eta_z<0$ region, while for the %higher density 
$R_s = R_{c1}$ transition, the $\hat{\bm \eta}_\parallel$ singularity jumps from the $\eta_z<0$ region into the $\eta_z \gtrsim 0$ region away from $\k=0$, nearly closing the quasiparticle energy gap. %We will see that %this is because while 
%of 
%We shall show 
This implies that the %two 
topological phase %transitions, %occurs at %the higher density transition at 
transition 
%both $R_s = R_{c1}$ and %involves only the change in topology while the lower density transition at 
%the one 
at $R_s = R_{c3}$ %the latter 
%will be shown to 
%also involves 
also involves the changes in %both topology and 
the exciton symmetry. %the Where as both  as it involves the exciton symmetry changing from the purely chiral $p$-wave to the near $s$-wave 
%whereas in the higher density (or smaller $R_s$) transition %with  %indicates that this 
%is nearly continuous %topological quantum phase transition %with the transition
%between s-wave dominant phase at high density and $p$-wave dominant phase at lower density, 
%has a much smaller discontinuity {\it i.e.}
%At the high density side of the transition point %at the point where 
%with the $s$-wave and the chiral $p$-wave components of $\langle \hat{\psi}^{\dagger}_{c,\k} \hat{\psi}_{v,\k} \rangle$ are nearly equal near this transition, which leads to 
%with 
%$|\Delta^p| \approx |\Delta^s|$ leads to a near gap closing away from the valley point ${\bf k}=0$.   
%Topologically, the low density transition is the inverse of the high density transition in the sense that the Berry phase reverts to the $\hat{H}_0$ value. 
%On the other hand, the lower density (or larger $R_s$) 
%this lower density transition is qualitatively more discontinuous as %in the sense that 
%the exciton symmetry changes from the purely chiral $p$-wave to the near $s$-wave. %, accompanied by the photon fraction jumping from zero to a finite value.  
%This transition %is probably coming from 
%originates from the competition within $\hat{H}^\text{MF}_\text{e-e}$ between the $s$-wave and the chiral $p$-wave symmetries when $\langle\hat{N}_{\text{ex}}\rangle$ becomes sufficiently small. 

\begin{figure}[h!]
	\includegraphics[height=115pt]{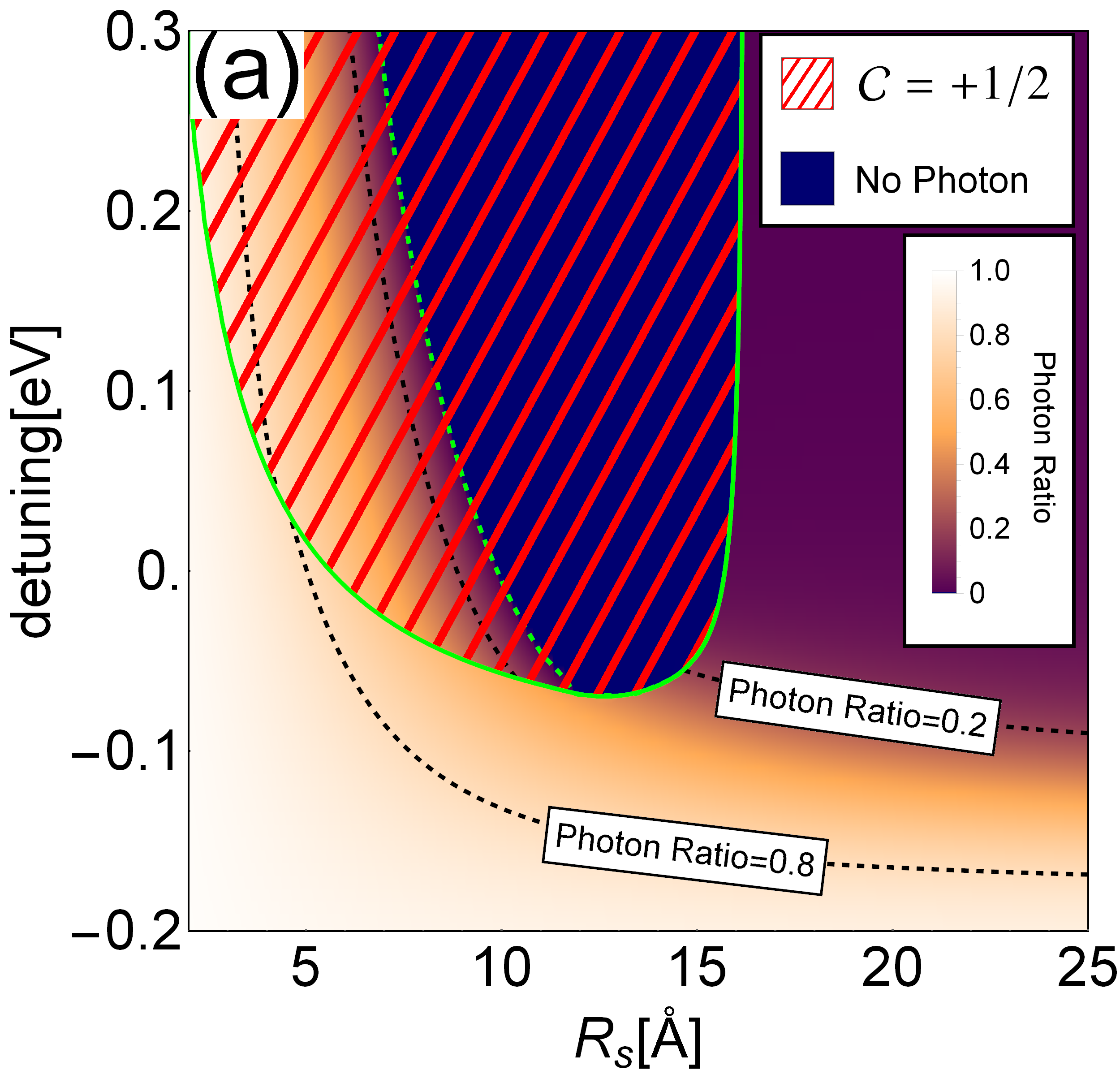}
	\includegraphics[height=112pt]{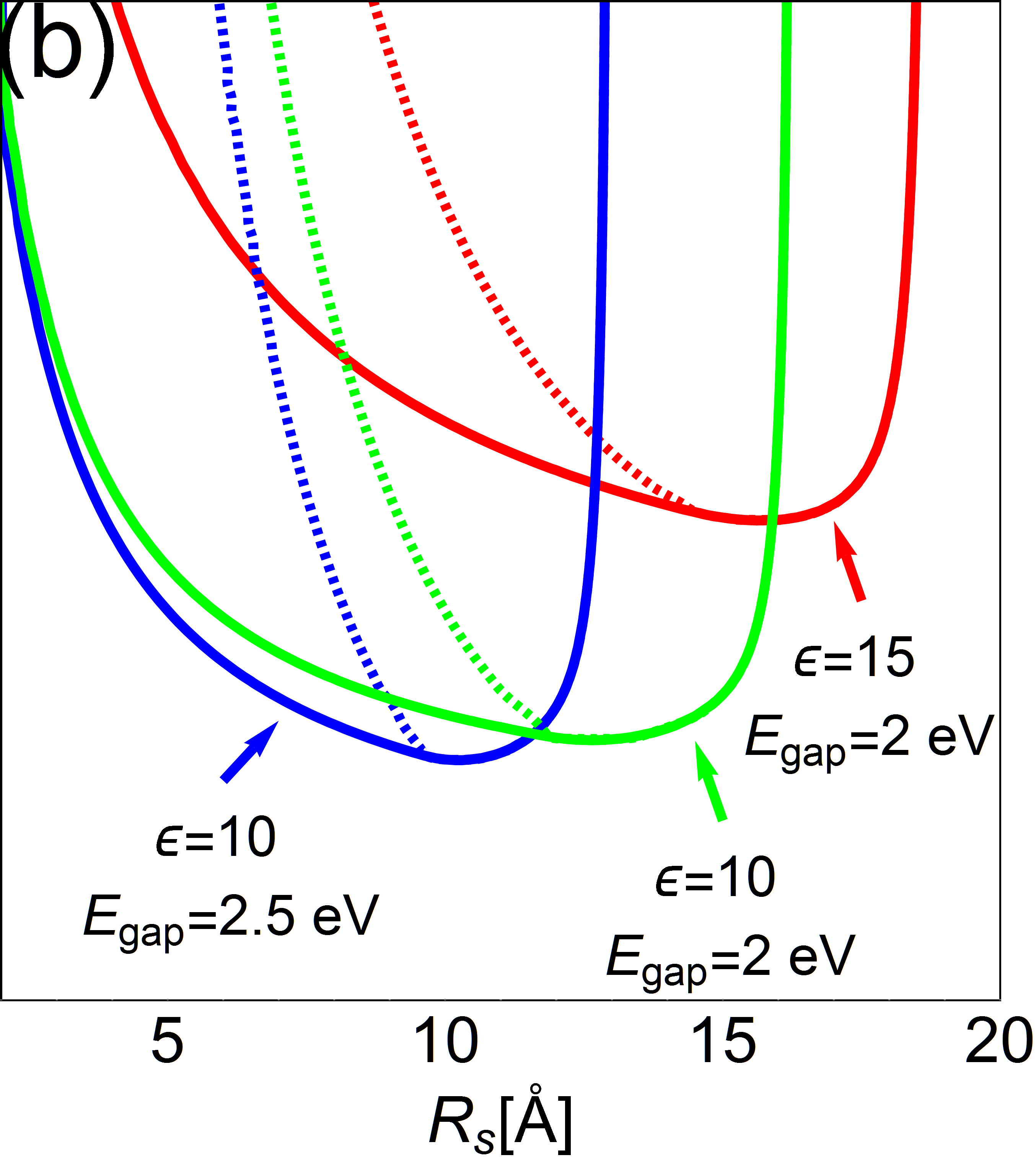}
	
	\caption{(a) The dependence of photon fraction for the single-valley TMDC polariton system on $\delta$ and $R_s$ shown with the velocity $\hbar v = \SI{3.7}{eV \AA}$, the dielectric constant $\epsilon = 10$ and $E_{\textrm{gap}}$= 2.0 eV; %, with (b) $\epsilon = 15$ and $E_{\textrm{gap}}$= 2.0 eV and (c) $\epsilon = 10$ and $E_{\textrm{gap}}$= 2.5 eV; $\hbar v=\SI{3.7}{\AA}$ is maintained for (a), (b), (c).  %It clearly shows that the $p$-wave exciton, which is dominant for $\mathcal{C}=+\pi$, is favored for the stronger Coulomb interaction (that is, lower $\epsilon$) and the smaller band gap. 
		%black curves represent the first-order transitions, green curves the second-order transitions, and dotted curves the crossovers. %The phases with crossovers have been discussed by Kamide {\it et al.}, the photon BEC, the polariton BEC and the exciton BEC. For $\mathcal{C}=+\pi$, we have the $p$-wave polariton BEC and the $p$-wave exciton BCS condensate.
		the green solid, the green dashed and the black dotted curves represent the first-order transitions, the second-order transitions, and the crossovers, respectively. 
		%The right plot shows how the $\mathcal{C} = +\frac{1}{2}$ region (full curves) and the second-order transitions (dotted curves) within the region change with $\epsilon$ and $E_{\textrm{gap}}$. %as well as the smallest detuning $\delta_c$. %showing the transitions are $-69.8$ meV, $38.9$ meV and $-79.78$ meV respectively.
		(b) Phase boundaries for the first-order (solid) and second-order (dashed) transitions for $\epsilon$=10 and $E_{\textrm{gap}}$=2 eV (green), $\epsilon$=15 and $E_{\textrm{gap}}$=2 eV (red), and $\epsilon$=10 and $E_{\textrm{gap}}$=2.5 eV (blue).
	}
	\label{FIG:phase2}
\end{figure}

%The nature of these phase transitions can be understood from the Berry curvature distribution taken at their vicinities, as shown in Fig. 2.
%The above considerations allow us to conclude 
Overall, the Fig.~\ref{FIG:phase1} plots %also shows us that the %phase transitions can be understood as the 
show how the topological phase transition of $\hat{H}^{\textrm{MF}}$ can arise from  the competition between the $s$-wave and the chiral $p$-wave exciton pairing channels. %as the 
%In Fig. \ref{FIG:phase1} (a), we have the Berry phase of $\mathcal{C} = \pi$ in the shaded region and $\mathcal{C}=-\pi$ in the unshaded region; note that %the switchings between $\mathcal{C}=-\pi$ and  
Note how the Chern number $\mathcal{C}_- = \pm\frac{1}{2}$ %($\mathcal{C}=-\frac{1}{2}$) 
coincides exactly with $|\Delta^s| < |\Delta^p|$ ($|\Delta^s| > |\Delta^p|$) in Fig.~\ref{FIG:phase1} (a). % occurs for %always coincide exactly with the crossings between  
%One way of obtaining the Berry phase is through 
$\mathcal{C}_-$ can be computed %from the pseudo-spin texture defined by $H^\textrm{MF}={\bm \sigma} \cdot {\bm \eta}(\k)$ %, the first-quantized form of $\hat{H}^{\textrm{MF}}$, through  
equivalently in either the orbital basis obtained from ${\bm \sigma} \cdot \hat{\bf d} = W ({\bf \sigma} \cdot \hat{\bm \eta}) W^\dagger$ or the band basis 
as $\mathcal{C}_-= \frac{1}{4\pi}\int d^2 k  \hat{\bf d} \cdot (\partial_{k_x} \hat{\bf d}  \times \partial_{k_y} \hat{\bf d}) 
= - \frac{1}{2} + \frac{1}{4\pi}\int d^2 k  \hat{\bm \eta} \cdot (\partial_{k_x} \hat{\bm \eta}  \times \partial_{k_y} \hat{\bm \eta})$, which is consistent with Fig.~\ref{FIG:phase1} (b)-(e) as it gives %the formula implies that 
$\mathcal{C}_- = \pm\frac{1}{2} $ %is obtained 
when %there is vorticity of the $x$, $y$ components of ${\bm \eta}$ 
the skyrmion %vorticity 
is present (absent); %for $\hat{\bm \eta}_\parallel$ in the $\eta_z > 0$ region. %Fig. \ref{FIG:phase1} (b)-(e) show graphically this change in $\mathcal{C}$ %the pseudo-spin texture at the phase transitions %is graphically shown in  the  , which show the pseudo-spin textures at the vicinities of the phase transitions, demonstrates graphically %that %these transitions occur when 
%the valley Berry phase %changes by $2\pi$ 
%reverses its sign from 
note that $\hat{H}_0$ gives $\mathcal{C}_- = -\frac{1}{2}$. %and the pseudo-spin in the orbital basis $\hat{\bf d}$ is related to that of the band basis by ${\bm \sigma} \cdot \hat{\bf d} = W ({\bf \sigma} \cdot \hat{\bm \eta}) W^\dagger$. %is obtained from the original band Hamiltonian . %value 
%this change  
%when 
%as $|\Delta^p| - |\Delta^s|$ changes its sign. %; note that we have .  
%Therefore the precise definition of 
In fact, we may define the overall exciton symmetry to be chiral $p$-wave %dominant phase a exciton is that 
when %the Berry phase of $\hat{H}^{\textrm{MF}}$ differs by $2\pi$ from that of $\hat{H}_0$. 
$\mathcal{C}_- = +\frac{1}{2}$. 
%Neither topological phase transitions of Fig. \ref{FIG:phase1} are continuous. 
Given that the $\Delta^p$ arises from the non-conservation of $N_{\rm tot}$ as can be seen from Eq.\eqref{EQ:exciton}, this is a case of discontinuous phase transitions to excitonic insulator phases in absence of the $N_{\rm tot}$ conservation, though our case deals with quantum %, or $T=0$, phase transitions 
rather than %the thermal 
classical phase transitions considered in %the previous studies 
\cite{keldysh_possible_1965, rice_theory_1973}. 

%This is further confirmed in the 
The full phase diagrams with respect to $R_s$ and the photon detuning $\delta \equiv \hbar\omega_c - E_{\textrm{gap}}$ shown as Fig. \ref{FIG:phase2} for different values of the dielectric constant $\epsilon$ and the band gap $E_{\textrm{gap}}$ can be largely explained 
by the different energy competitions that give rise to the higher and the lower density phase transition. %While we find the phases obtained in the conventional polariton condensate whose boundaries represent crossovers, we also 
%The point here is 
%The photon detuning serves as a 
$\delta$ and $\epsilon$ are control parameters in the competition between the e-ph coupling and the e-e interaction; the photon self-consistency equation Eq.\eqref{eq:sceq_lambda} shows that the smaller $\delta$ leads to the larger photon fraction, while the smaller $\epsilon$ leads to the larger e-e interaction. Fig.~\ref{FIG:phase2} (b) shows that the $\mathcal{C}_- = +\frac{1}{2}$ phase with the chiral $p$-wave excitons requires %$\delta > \delta_c$, %(about 40meV for $\epsilon = 15$, $E_g = 2$eV), 
%{\it i.e.} a 
sufficiently weak %electron-photon 
e-ph coupling, which is naturally larger for the smaller Coulomb interaction of $\epsilon = 15$ shown in %Fig. \ref{FIG:phase2} (b) 
red than for the larger Coulomb interaction of $\epsilon = 10$ shown in %Fig. \ref{FIG:phase2} (a) and (c).  
blue and green. %For $\delta > \delta_c$, all Fig. \ref{FIG:phase2} plots 
That %the right plot shows %dependence on $\delta$ to be significantly stronger for 
the lower density (larger $R_s$) transition depends little on $\delta$ %compared to the higher density (smaller $R_s$) transition, 
%which is consistent with %only  
confirms %the lower density (larger $R_s$) transition depending 
its weak dependence on the 
%We therefore conclude from 
%Therefore Fig 2 that %the strong dependence of 
%the higher density (smaller $R_s$) transition %on $d$ is another evidence that it is a direct consequence of 
%arising from the competition between the %electron-electron 
%e-e interaction and the %electron-photon 
e-ph coupling. %while %the weak dependence of 
%the former %on $d$ indicates that it 
%having little to do with the electron-photon coupling. 
%When we compare Fig. \ref{FIG:phase2} (a) with (b) and (c), the  $\mathcal{C} = \pi$ region is shown to shrink as expected when the increased $\epsilon$ decreases the Coulomb interaction and when 
Meanwhile, the blue curves of Fig.~\ref{FIG:phase2} (b) shows that for a larger $E_{\rm gap}$ the lower density transition occurring at smaller $R_s$ (larger density) when compared with the lower $E_{\rm gap}$ shown %in Fig.\ref{FIG:phase2} (a) and (b). %, a consequence, %This is because, 
by the green and red curves. This is because of the larger $E_{\rm gap}$ suppressing $\Delta^p$ through reducing $\sin \theta_k$ at all momenta, or, equivalently, the Berry curvature integrated over momenta smaller than $k$. %, hence suppressing $\Delta^p$, and 
%as can be seen from Eq.\eqref{EQ:exciton}, %that 
%of a larger $E_{\rm gap}$ %effectively expands the $\k$-space region in vicinity of the valley point, leaving %have shown that at low photon density, $\Delta^s (k)$ with its overall factor of $\cos^2 \theta_{\bf k}/2 = (1+E_{\rm gap}/\sqrt{E^2_{\rm gap}/4+(\hbar v_F k)^2})/2$ is favored over $\Delta^p (k)$ with its overall factor of $\sin \theta_{\bf k} = h v_F k / \sqrt{E^2_{\rm gap}/4+(\hbar v_F k)^2}$ when $E_{\rm gap}$ is increased; this 
%can also be interpreted as the result of 
%leaving 
%the Berry curvature being less concentrated at the valley point. %; %with the increased $E_{\rm gap}$; 
%this favors $\Delta^s$ over $\Delta^p$, which is induced by the valley Berry phase.

%(a paragraph discussing on the rotational symmetry breaking for excitons with the mixed $s$-wave and $p$-wave symmetries)

%Overall, 
The phase diagram of Fig.~\ref{FIG:phase2} (a) shows phase transitions as well as crossovers in contrast to the results for the polariton condensate in the topologically trivial quantum well where only the latter were present \cite{kamide_meanfield_2010}. 
Following the results of Kamide {\it et al.} for the topologically trivial quantum well, we can define in the $\mathcal{C}_- = -\frac{1}{2}$ region several phases %just as in the topologically trivial quantum well, distinguished by 
according to the photon fraction %or the exciton Bohr radius compared to $R_s$; 
%between these phases - which  labeled, with decreasing photon fraction, 
as the photon, the polariton and the exciton BEC in the decreasing order, with their boundaries being crossovers (shown as the dotted curves). However, as discussed above, there is a first-order phase transition (shown as the solid curves) %at the boundary 
between the $\mathcal{C}_- = -\frac{1}{2}$ and the $\mathcal{C}_- = +\frac{1}{2}$ regions. Within the $\mathcal{C}_- = +\frac{1}{2}$ region, the phase with the vanishing photon fraction would be best termed the electron-hole BCS condensate, %where BCS refers 
with BCS %referring to 
indicating the %$p$-wave 
%excitons having a 
exciton radius being larger than $R_s$. %Because of this, for the energy of the $p$-wave exciton condensate, the interaction between excitons is more relevant than the binding energy of each exciton, which is the main reason why the $p$-wave exciton condensate can be stable when each $p$-wave exciton always has higher energy than the $s$-wave excitons. 
Inside the $\mathcal{C}_- = +\frac{1}{2}$ region, there is %phase 
%transition from the polariton BEC to this electron-hole BCS condensate is %should be of 
a second-order phase transition (shown as the dashed curves) %, being continuous and 
between the polariton BEC and this electron-hole BCS condensate
involving %due to 
the spontaneous rotational symmetry breaking. %that we shall now explain.
%For the one-valley problem, spontaneous rotational symmetry breaking generically occurs for the polariton condensation with circularly polarized photons in TMDC cavity. %requires phase transition even when the Berry phase is unchanged from that of  $\hat{H}_0$ due to the 
%This is due to 
Despite %there being no 
%absence of 
photons providing %the 
no preferred direction, %in the circular photon polarization, 
the rotational symmetry is broken %in the presence of 
when we have both the $s$-wave and the chiral $p$-wave components in $v_{\bf k}/u_{\bf k}$ of the exciton wave function Eq.\eqref{EQ:WF}, %, except when the photon fraction vanishes completely as in Fig. \ref{FIG:phase1} (d). 
%This is why 
%and can be seen from 
%one of its consequence being 
which moves 
%moving 
the singularity of ${\bm \eta}_\parallel$ textures of Fig. \ref{FIG:phase1} (b), (c), (e) %are 
%not centered at 
away from %the 
$\k=0$. %valley point. %, the minimum of the exciton amplitude being in the direction of this texture center. %, thereby breaking the rotational symmetry around the valley point. 
%This is a spontaneous symmetry breaking as the circularly polarized photon is not biased in any direction. %It persists for more realistic $\hat{H}_0$ where higher order corrections give us the threefold rather than the continuous rotational symmetry, %as that would only change 
%merely modifying the spontaneous breaking of a continuous symmetry to that of a discrete symmetry. 
%It is possible for $\Lambda$ and $\Delta^s$ to vanish in a continuous phase transition only because this vanishing leads to the restoration of the rotational symmetry in our polariton condensation, %. %, there needs to be a continuous phase transition from the non-interacting state %in spite of inclusion in our analysis of the Coulomb interaction terms Eq.\eqref{EQ:exciton} that does not 
%despite the non-conservation of the exciton number $N_{\rm ex}$ in our analysis as shown in Eq.\eqref{EQ:exciton}. %given by $\hat{H}_0$ even when the Berry phase remains $\mathcal{C} = -\pi$ as in $\hat{H}_0$.
%This is because 
The rotational symmetry in our polariton condensate is restored in Fig.~\ref{FIG:phase1} (d) on $\Lambda$ and $\Delta^s$ vanishing continuously. %, the vanishing photon fraction giving us a purely chiral $p$-wave $v_{\bf k}/u_{\bf k}$, the only case when our condensation does not involve the rotational symmetry breaking. %is when , for which, , $\mathcal{C} = \pi$, which implies second-order phase transition from Fig. \ref{FIG:phase1} (d) to Fig. \ref{FIG:phase1} (c) involving spontaneous rotational symmetry breaking. 
%Lastly we note that the comparison between Fig. \ref{FIG:phase1} (d) and (e) shows that the lower density topological phase transition also involves symmetry breaking as well. 
Hence %our polariton condensate has 
%either the topology or the rotational symmetry of 
our polariton condensate always possesses either topology or symmetry distinct from the ground state of $\hat{H}_0$. 

\begin{figure}
\centering{\includegraphics[width=\columnwidth]{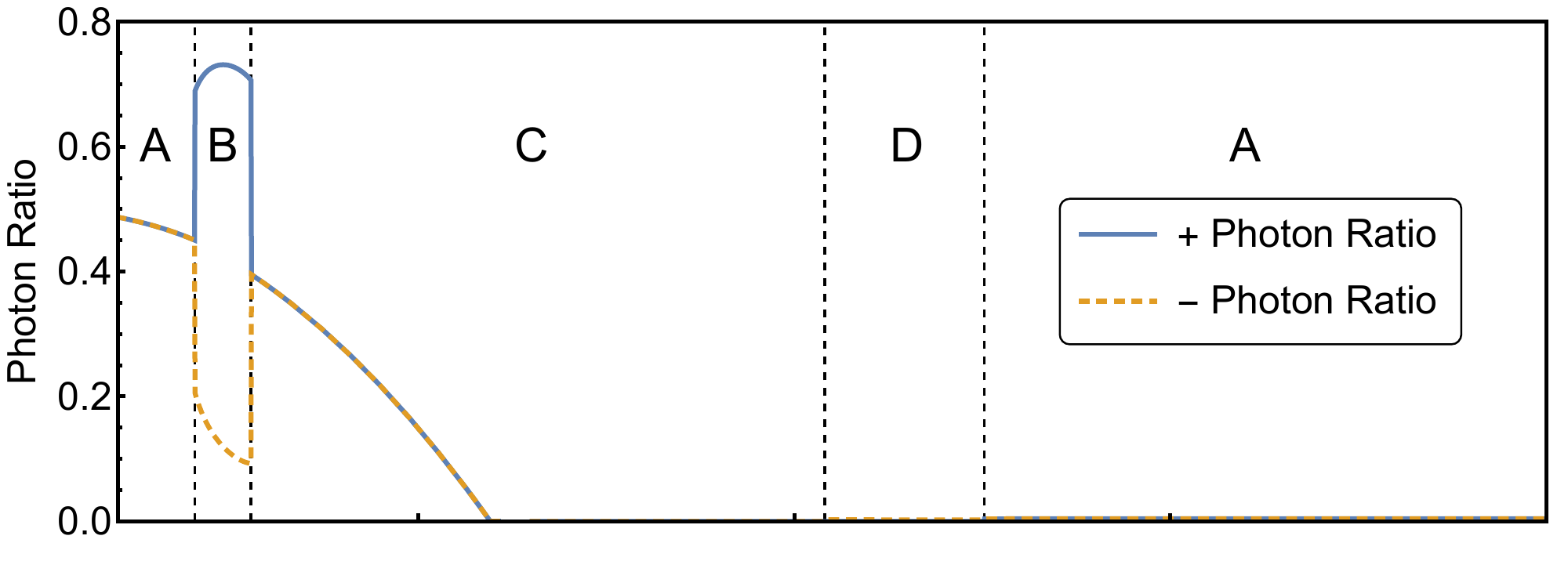}}\vspace{-.5em}
\centering{\includegraphics[width=\columnwidth]{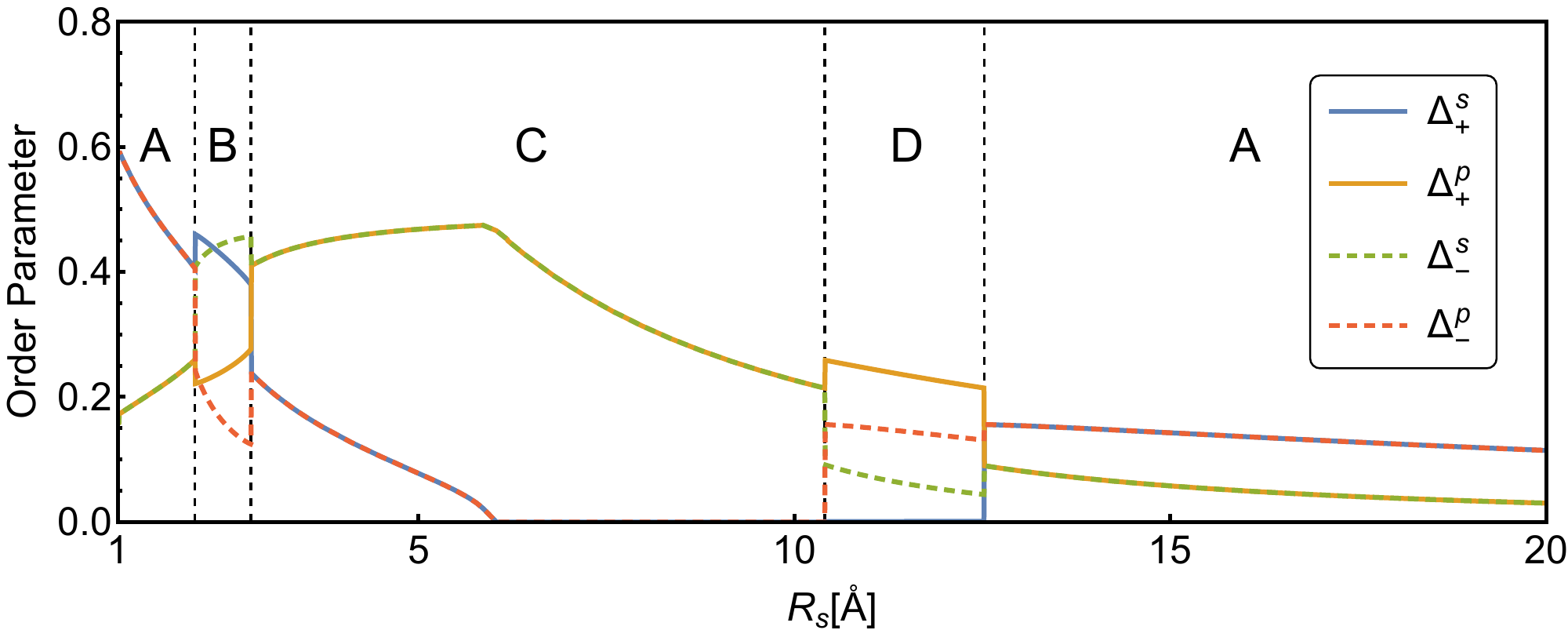}}
	
	\caption{Photon fraction (above) and mean-field band exciton gap parameters $\Delta^{s,p}$ (below) for two valleys ($\tau = \pm 1$) %and $\mu_X$ 
	as the functions of $R_s$ for the photon frequency $\hbar\omega_c = \SI{2.1}{eV}$ and other physical parameters following those of Fig. \ref{FIG:phase2} (a). %the dielectric constant $\epsilon = 10$, the Dirac velocity $\hbar v = \SI{3.7}{eV \AA}$, and the band gap of $E_{\textrm{gap}} = \SI{2.0}{eV}$ .
	}	

	\label{FIG:twovalley}

\end{figure}

%The topological phase transitions arising from the Berry phase sign change can be well-defined only when %For the experimental detection of the $\pi$ Berry phase effect in the Dirac material polariton condensate, we need to consider the phase transitions for the case where 
%we include both valleys in our analysis. 
%The phase transition of the polariton condensate in the two valley TMDC largely follows that of the one valley problem we have discussed above. %, although %Next, we put together the mean-field Berry phase at each valley to obtain 
%elucidating the topological phase of TMDC polariton condensate does require the Berry phase of $\hat{H}_{MF}$ from both valleys. 
For the two valley TMDC %with the linearly polarized initial laser pulse, 
coupled to photons of both circular polarizations shown in Fig. \ref{FIG:twovalley}, we find that the topological phase transitions give rise to both the quantum spin Hall phase (in the region C) and the quantum anomalous Hall phase (in the regions B and D). %spontaneous valley polarization occurs for two %separate ranges of $R_s$. 
%regions denoted as B and D. 
%For this two-valley problem, 
To analyze this problem, we consider the variational solution of Eq. \eqref{EQ:WF} %not only with both valleys and photon polarizations ${\bf e}_{\pm} = (1,\pm i)/\sqrt{2}$, but also 
with the phase difference between the two photon polarizations fixed. %, %; this 
%which would correspond physically to the linearly polarized initial laser pulse. 
%As shown in Fig. \ref{FIG:twovalley}, we find valley polarized regime, marked by B and D, near the first-order phase transition points for single valley problem. %in terms of the mean distance between excitations of the same valley, the phase transitions %of the two-valley solution occur near the point where the Chern number sign reversal occurred in 
%each valley 
%at the two valleys occur quite close to that of the one-valley solution of Fig. \ref{FIG:phase1} (a). %; in addition, %when the Berry phase sign is reversed, 
%the vanishing of the photon fraction persists for the corresponding range of $R_s$. %Both remain true even when we modify Eq.\eqref{eq:total Hamiltonian} with the more realistic $\hat{H}_0$ with higher order corrections that allow for the inter-valley coupling through photons. 
%However, the %sign change of $|\Delta^p|-|\Delta^s|$ 
%first-order topological phase transition points do not quite coincide between the two valleys, resulting in the valley polarization when we have $|\Delta^p_\tau|>|\Delta^s_\tau|$ but $|\Delta^p_{\bar \tau}|<|\Delta^s_{\bar \tau}|$ in the regions B and D of Fig. \ref{FIG:twovalley}. %Given that , when 
In the absence of interactions, $\hat{H}_0$ of Eq. \eqref{eq:total Hamiltonian} gives us the opposite sign for the Chern numbers of $\mathcal{C}_\tau = \frac{\tau}{2}$ for the $\tau$ valley. 
When the exciton symmetry of one valley is the chiral $p$-wave and that of the other valley is the $s$-wave, we have a net Chern number of $\mathcal{C}_{\rm tot}\equiv\sum_{\tau,\sigma} \mathcal{C}_{\tau,\sigma} = \pm 1$ %, which means 
for our $\hat{H}^{\rm MF}$ and hence %is in 
the quantum anomalous Hall phase \cite{haldane_hall_1988}. %when we count in the contributions from the bands away from the Fermi level as well. 
%To consider 
%Then, the photoluminescence 
Due to the valley polarization that occurs only in this phase, %the photon polarization in 
the regions B and D have the elliptic photon polarizations %, instead of linear, photon polarization for this phase. 
%rather than linear as in 
while all the other regions have the linear photon polarizations. 
%Yet this does not necessarily imply a strong coupling between the two valleys; for instance,  as shown in the region C of 
Meanwhile, the region C of Fig. \ref{FIG:twovalley} shows that %the continuous vanishing of 
the photon fraction and the $\Delta^s$ at both valleys %in the region C %vanish  %continuously 
%in the quantum spin Hall phase 
vanish  continuously at the same $R_s$ %and this vanishing is continuous despite the rotational symmetry breaking no longer being spontaneous with the linearly polarized photons. %and moreover is continuous just as they are for the $\Delta^p$-dominant regime in Fig. \ref{FIG:phase1} (a). 
\footnote{The rotational symmetry breaking at two valleys are not independent due to the e-ph coupling. Therefore, if we take the $\hat{H}_0$ of Eq.\eqref{eq:total Hamiltonian}, {\it i.e.} with the continuous rotational symmetry, we have for the two valley case the SO(2) rather than SO(2) $\times$ SO(2) symmetry breaking. Given the photon polarization, the vanishing photon fraction is necessary for the rotational symmetry.}. %; this mode is gapped once we reduce the rotational symmetry of $\hat{H}_0$ to the threefold symmetry.}  
In the region C, we have the quantum spin Hall phase %in the region C %of Fig. \ref{FIG:twovalley} %, as the time-reversal symmetry is restored there.  
where the time-reversal symmetry is restored by the opposite chirality between the $p$-wave excitons of the two valleys. 
%For $|\Delta^p_\tau|>|\Delta^s_\tau|$ in both valleys, time-reversal invariance is restored in the TMDC polariton condensate, giving us the quantum spin Hall phase for $\hat{H}^{\rm MF}$. %Meanwhile, the quantum spin Hall is the topological phase %for the case where 
%obtained when the Berry phase sign is reversed at each valley, %we need to note that 
%This is mainly because 
%This occurs because %We have seen from Eq.~\eqref{EQ:valleyChirality} that 
%The time-reversal symmetry required for this phase is restored both due to the $p$-wave excitons in the two valleys having the opposite chirality as obtained in Eq.~\eqref{EQ:valleyChirality} and also the linear photon polarization. %which can restore the time-reversal symmetry of the TMDC polariton condensate %can be restored 
%(hence the zero net Chern number). %along with the equal exciton amplitude for two valleys. %therefore giving us the zero net Chern number. 
%In order to properly determine 
Table 1 shows the topological phases for the two-valley TMDC polariton condensate %as shown in requires %we need to %note that for each valley in 
%going beyond the effective spin-polarized valley picture of Eq.\eqref{eq:total Hamiltonian} %is effectively spin polarized to have the opposite spin at the band edge due to the atomic spin-orbit coupling. Taking 
%and 
taking into account %the other 
both spin components at each valley. %for which %is little affected by 

\begin{table}
	\begin{tabular}{|c|c|c|c|c|c|c|c|}
		\hline
		& $\mathcal{C}_{+\uparrow} $ & $\mathcal{C}_{+\downarrow} $  & $\mathcal{C}_{-\uparrow}$ &  $\mathcal{C}_{-\downarrow}$ & $\mathcal{C}_{\rm S}$ & $\mathcal{C}_{\rm V}$ & $\mathcal{C}_{\rm tot}$ \\
		\hline
		A &     +1/2 &  +1/2 & $-$1/2 &     $-$1/2 &    0 &   +1 &      0 \\
		B & $\pm$1/2 &  +1/2 & $-$1/2 & $\pm$1/2 & $-$1/2 & +1/2 & $\pm$1 \\
		C &     $-$1/2 &  +1/2 & $-$1/2 &     +1/2 &   $-$1 &    0 &      0 \\
		D & $\mp$1/2 &  +1/2 & $-$1/2 & $\mp$1/2 & $-$1/2 & +1/2 & $\mp$1 \\
		\hline
	\end{tabular}
	\caption{Phase classification in the two valleys %case with the linear photon polarization. 
		coupled to the photons of both circular polarizations. The alphabet letters in the leftmost column refer to each phase mentioned in Fig. \ref{FIG:twovalley}. %SH stands for spin Hall, VH stands for valley and AQH stands for anomalous quantum Hall. Note that SH and VH is only an approximation as both spin and valley are not strictly conserved. 
		$\mathcal{C}_{\rm S} \equiv \sum_{\tau,\sigma} \sigma\mathcal{C}_{\tau,\sigma}/2$ and $\mathcal{C}_{\rm V} \equiv \sum_{\tau,\sigma} \tau\mathcal{C}_{\tau,\sigma}/2$ are the spin and the valley Chern numbers respectively. Refer the main text for further details.}
\end{table}

In summary, we have studied the quasi-equilibrium ground state of the TMDC monolayer coupled to the cavity photons with the self-consistent mean-field theory and found topological phase transitions due to the competition between the e-ph coupling and the e-e interaction tuned by the excitation density. %We have not made any attempts to consider the dynamics of our system, including any 
%To be consistent with this approach, we have not considered in our analysis the non-zero photon loss rate, which may lead the system to the lasing state rather than the polariton condensate, as it represents a characteristic of the cavity rather than TMDC. 
Our approach is expected to work best %in the limit of the infinite polariton lifetime. %however, it does need to be mentioned that, as our process begins with the initial photon pulse optically creating the $s$-wave excitons, the equilibration time may %be lengthened if the $p$-wave 
%depend on the relative magnitude of the $s$-wave component of our excitons %dominate 
%at the quasi-equilibrium. 
for the thermal quasi-equilibration time shorter than the polariton lifetime. %at all range of $R_s$. 
%In reality, this quasi-equilibration time may have dependence on various physical parameters, {\it e.g.} $R_s$, %; 
%Consequently, we can not rule out our phase diagram being partially 
%Due to such dependence, 
%which would make certain regions of our phase diagram more accessible experimentally than others. %due to the possible dependence of both the quasi-equilibration time and the polariton lifetime on $R_s$.%While the possibility of electron-hole plasma has been discussed for the indirect exciton condensate in the bilayer quantum well, our system lacks the energetics argument for this state to occur, namely the competition between the intralayer and interlayer Coulomb interaction.
We may find the regions of our phase diagram with optimal experimental accessibility %given the possible dependence of 
as the quasi-equilibration time may depend on various physical parameters, {\it e.g.} $R_s$. One possible method for triggering our phase transitions may be the terahertz pump which has been shown to induce the $s$-wave to $p$-wave transition in the excitons \cite{Menard_THzPump_2014}.

\begin{acknowledgments}
{\it Acknowledgement}: We would like to thank Xi Dai, Hans Hansson, Joon Ik Jang, Na Young Kim, S. Raghu, Yao Wang and Fengcheng Wu for sharing their insights. S.B.C. %would like to thank 
acknowledges the hospitality of Nordita during its conference on ``Multi-Component and Strongly Correlated Superconductors" where parts of this work has been completed.  This research was supported by IBS-R009-Y1 (K.H.L. and S.B.C.) and Basic Science Research Program through the National Research Foundation of Korea (NRF) funded by the Ministry of Education under Grant No. 2015R1D1A1A01058071 (C.L. and H.M.).
\end{acknowledgments}

\bibliography{ref}

\end{document}